\documentclass[%
 reprint,
superscriptaddress,
 amsmath,amssymb,
 aps,
prb,
]{revtex4-2}

\usepackage{graphicx}
\usepackage{dcolumn}
\usepackage{bm}
\usepackage{xcolor}
\usepackage{soul}
 
\usepackage[unicode]{hyperref}
\hypersetup{
   plainpages=false,
   colorlinks=true,       
   citecolor=blue,        
}
\usepackage{orcidlink}

\begin{document}
\renewcommand{\appendixname}{APPENDIX}

\title{
Quartic level repulsion in a quantum chaotic three-body system without symplectic symmetry
}

\author{Alex D. Kerin
\orcidlink{0000-0001-9204-9116}}
\email{Contact: dradkerin@gmail.com}
\affiliation{Te Whai Ao Dodd-Walls Centre for Photonic and Quantum Technologies, Auckland 0745, New Zealand} 
\affiliation{Centre for Theoretical Chemistry and Physics, New Zealand Institute for Advanced Studies, Massey University, Private Bag 102904, North Shore, Auckland 0745, New Zealand}
\author{Barbara Dietz \orcidlink{0000-0002-8251-6531}}
\email{Contact: bdietzp@gmail.com}
\affiliation{Center for Theoretical Physics of Complex Systems, Institute for Basic Science (IBS), Daejeon 34126, Republic of Korea}
\author{Joachim Brand \orcidlink{0000-0001-7773-6292}}
\email{Contact: j.brand@massey.ac.nz}
\affiliation{Te Whai Ao Dodd-Walls Centre for Photonic and Quantum Technologies, Auckland 0745, New Zealand} 
\affiliation{Centre for Theoretical Chemistry and Physics, New Zealand Institute for Advanced Studies, Massey University, Private Bag 102904, North Shore, Auckland 0745, New Zealand}

\date{\today}

\begin{abstract}
Among the fundamental symmetry classes of quantum chaotic systems in Dyson's threefold way, the symplectic class is rarely observed in nature. Characterized by the strongest possible level repulsion in the energy spectrum, the symplectic symmetry class also implies a double (Kramers) degeneracy of levels. Studying the spectral statistics of three quantum particles (identical bosons or mass-imbalanced fermions) in a harmonic trap, we find numerical evidence for strong level repulsion in the regime of weak contact interactions. While the statistical indicators are consistent with quantum chaos in systems with symplectic symmetry, the absence of Kramers degeneracy rules out this symmetry. In the strongly-interacting unitary limit either Poissonian or stick statistics are observed (depending on commensurability of the mass ratio) indicating regular dynamics. 
The transition between the regular and chaotic regimes as a function of interaction strength is well described by the Rosenzweig-Porter model. 
The system can be realized with cold neutral atoms in microtraps or in optical lattices.
\end{abstract}

\maketitle

\section{Introduction} Random matrix theory (RMT) has been a very useful tool in extending the classical concept of chaos to the quantum realm. According to the Bohigas-Gianonni-Schmit conjecture (BGS)~\cite{Bohigas1984}, the statistical properties of a typical Hamiltonian's eigenspectrum are well described by RMT if the underlying classical dynamics are chaotic~\cite{wigner1993characteristic,casati1980connection, Haake2018, Stockmann1999a}. In typical quantum systems with regular dynamics the energy levels of a given symmetry subspace are uncorrelated, and the nearest-neighbor level spacings, $S$, are distributed as a Poissonian, $P(S)=\exp(-S)$. 
For quantum systems with chaotic dynamics adjacent levels repel, and $P(S)$  is well approximated by the Wigner surmise $P(S)\propto S^\beta \exp(-A_\beta S^2)$ where $A_{\beta}$ is a constant, and $\beta=1,2$ or $4$ depending on the symmetries of the particular symmetry subspace in question. The chaotic or regular nature of a quantum system has significant physical impact. It informs how the system equilibrates~\cite{Deutsch1991,Srednicki1994, torres2015dynamics, Alessio2016, torres2017dynamical, torres2017extended, torres2018generic, santos2018nonequilibrium, schiulaz2019thouless, de2020quantum}, and is deeply tied to questions of localization and ergodicity~\cite{Pal2010, Serbyn2013, Kravtsov2015, Pino2019,alet2018many, vsuntajs2020quantum, bulchandani2022onset, santhanam2022quantum,abanin2017recent,abanin2019colloquium}.

The usually assumed ergodic dynamics of a gas of neutral atoms is important for the success of laser and evaporative cooling techniques \cite{Cohen-Tannoudji2011a, phillips1982laser, chu1991laser, tannoudji1992atom}, and gives rise to loss rates due to three-body recombination \cite{Burt1997,Esry1999a, eismann2016universal}. A recent experiment observing much reduced three-body losses in individual atomic triads \cite{reynolds2020direct} has highlighted the fact that quantum chaos in trapped atomic few-particle systems is poorly understood.
Previous theoretical studies on the level statistics of trapped interacting gases were either limited to one-dimensional systems~\cite{Kolovsky2004a,Chakrabarti2012, Roy2012, huber2021morphology, fogarty2021probing, lydzba2022signatures, AnhTrai2023QuantumChaos, de2024thermalization}, or considered the many-body case~\cite{Chakrabarti2012,Roy2012}.



In this work we consider a quantum gas of three interacting bodies in three dimensions and search for signatures of chaos in the spectral statistics. Specifically, we consider three contact-interacting quantum particles in a spherical harmonic trap, modeling the dynamics of cold neutral atoms. We utilize semi-analytic and numerical techniques that allow for accurate calculation of the energy spectrum as a function of the scattering length $a_s$, which parametrizes the strength of the contact interaction~\cite{kestner2007level, liu2009virial, liu2010three, werner2006unitary, Werner2006unitarygas, werner2008trapped, kerin2023energetics}. 
The strongly-interacting limit, $a_s\to\infty$, is a special case known as the unitary limit because the cross section is maximized under the constraint of a unitary scattering process \cite{braaten2013universal}. In this case,
explicit solutions of the three-body problem are known and we find either Poissonian or clustered, ``stick'', statistics depending on a commensurability condition for the mass ratio. The non-Poissonian stick statistics are reminiscent of the (non-interacting) harmonic oscillator, a paradigmatic example of an atypical integrable system not following the BGS conjecture~\cite{Berry1977,Drod1991, Chakrabarti2003}. For finite interaction strength we observe a continuous transition between regular behavior for strong interactions and chaotic behavior for weak interactions. Curiously, we observe strong level repulsion and other statistical signatures consistent with $\beta=4$. This is unusual because the time-reversible and spinless nature of this system implies $\beta=1$~\cite{Haake2018}. The system's other symmetries may be responsible for this result~\cite{joyner2014gse}.

\section{The model} We consider three interacting particles. Either two identical fermions plus an impurity or three identical bosons in a spherical harmonic trap with zero-range interactions. The Hamiltonian of this system is the simple harmonic oscillator Hamiltonian
\begin{eqnarray}
H=\sum_{i=1}^{3}\frac{-\hbar^2}{2m_{i}}\nabla_{i}^2+\frac{m_{i}\omega^2 r_{i}^2}{2},\label{eq:Hamiltonian}
\end{eqnarray}
where $m_{i}$ and $r_{i}=\vert \vec{r}_{i} \vert$ are the $i^{\rm th}$ particle's mass and position respectively, with $\omega$ being the trapping frequency. We do not consider the spin of the particles beyond ensuring the appropriate bosonic/fermionic exchange symmetries; cf. the appendix. The interactions between particles $i$ and $j$ are enforced by the Bethe-Peierls boundary condition~\cite{bethe1935quantum}
\begin{eqnarray}
\lim_{r_{ij}\to 0}\left[ \frac{1}{r_{ij}\Psi} \frac{d}{dr_{ij}}(r_{ij}\Psi)\right]=\frac{-1}{a_{s}}, \quad r_{ij}=|\vec{r}_{i}-\vec{r}_{j}|,\label{eq:BethePeierls}
\end{eqnarray}
where the wave function $\Psi$ depends on the coordinates of all particles $\{r_i\}$. While the $s$-wave scattering length $a_{s}$ becomes equivalent to a hard-sphere radius in a low-density and low-energy limit if $a_s\ge0$, with $a_{s}=0$ the non-interacting case, there is no general classical limit of the contact interaction model.
However, systems with no obvious classical analogue have been shown to follow RMT e.g. nuclear systems~\cite{Guhr1989,Zelev1996,guhr1998random,Weidenmueller2009,giasemis2022quantum, villasenor2024breakdown,Dietz2017}.

The energy spectrum of this system is known semi-analytically at unitarity, $a_s\to\infty$,~\cite{werner2006unitary, Werner2006unitarygas, werner2008trapped} and numerically for arbitrary $a_s$ to great accuracy~\cite{kestner2007level, liu2009virial, liu2010three}. The center-of-mass motion is a non-interacting simple harmonic oscillator, and we do not consider it further. The relative motion contains eigenstates that are unaffected by the interactions (the laughlinian states \cite{Werner2006unitarygas,Endo2016,Bradly2025}) and states that are affected. For simplicity we focus on the interacting spectrum with zero angular momentum. 

\begin{figure}[h]
\includegraphics[width=8cm, height=5cm]{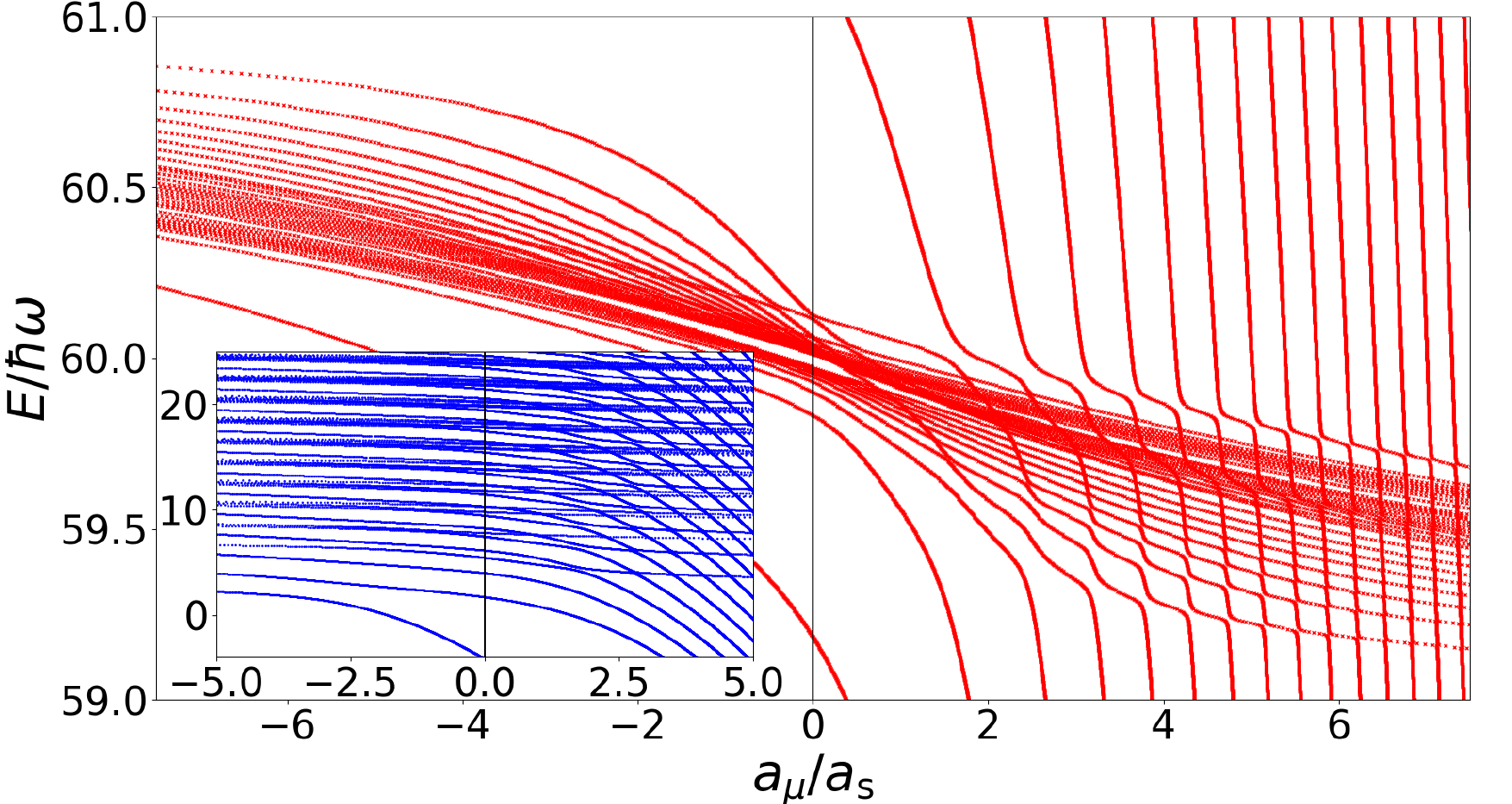}
\caption{ The relative-motion energy spectrum defined by Eqs.~\eqref{eq:Hamiltonian} \& \eqref{eq:BethePeierls} for fermions with an equal mass impurity and $l=0$. Inset: The relative energy spectrum for bosons with $l=0$. The spectrum qualitatively is similar for different particle symmetries or impurity masses. $a_{\mu} = \sqrt{\hbar/\mu\omega}$ is the simple harmonic oscillator length scale of the relative motion.}
\label{fig:ESpec}
\end{figure}

We refer the reader to the appendix for a brief review of how the energy spectrum is calculated. In Fig.~\ref{fig:ESpec} we plot the energy spectrum. The spectrum is organized into ``clusters'' of states spaced by $\approx 2\hbar\omega$. Notably, avoided-crossings are visible for $a_s>0$. These indicate the presence of level repulsion, and that we have isolated a single symmetry subspace.

\section{Unfolding the spectrum} In order to sensibly compare a physical energy spectrum to the predictions of RMT the spectra must be rescaled (or ``unfolded'')~\cite{guhr1998random, gomez2002misleading, abul2014unfolding, santos2018nonequilibrium, abuelenin2018spectral} to remove secular variations in level density. The size of fluctuations in level spacing must be compared to the local scale of level separation. One common way to unfold is to numerically fit a function, $N^{\rm smooth}(E)$, to the cumulative level density $N(E)$, the number of levels with energy less then $E$, such that $N(E)=N^{\rm smooth}(E)+N^{\rm fluc}(E)$, where $N^{\rm fluc}$ is a small fluctuating component that is zero on average. The unfolded energies are $\epsilon_{i}=N^{\rm smooth}(E_{i})$, where $E_i$ are the original energy eigenvalues, and the level spacings are $S_{i}=\epsilon_{i+1}-\epsilon_{i}$. By construction $\langle S \rangle =1$. Here the spectrum is divided into distinct clusters of states. We unfold each cluster individually by fitting a function of the form
\begin{eqnarray}
N^{\rm smooth}(E)=\begin{cases}
\sum_{n=1}^{n_{\rm max}}a_n(C-E)^{-n}, \quad E>C\\
\sum_{n=1}^{n_{\rm max}}b_n(C-E)^{-n}, \quad E<C\\
\end{cases},
\label{eq:Nfit}
\end{eqnarray}
where $C$ is a constant denoting the central point of the cluster, and $a_n$ and $b_n$ are fitting parameters. The integrated spectral density $N(E)$ for the whole spectrum can be obtained by concatenating the fits for individual clusters. In the below, we use $n_{\rm max}=2$ for finite $a_s$ and $n_{\rm max}=4$ for unitarity.

\section{Level-spacing statistics}
\begin{figure*}
    \centering
    \includegraphics[width=16cm]{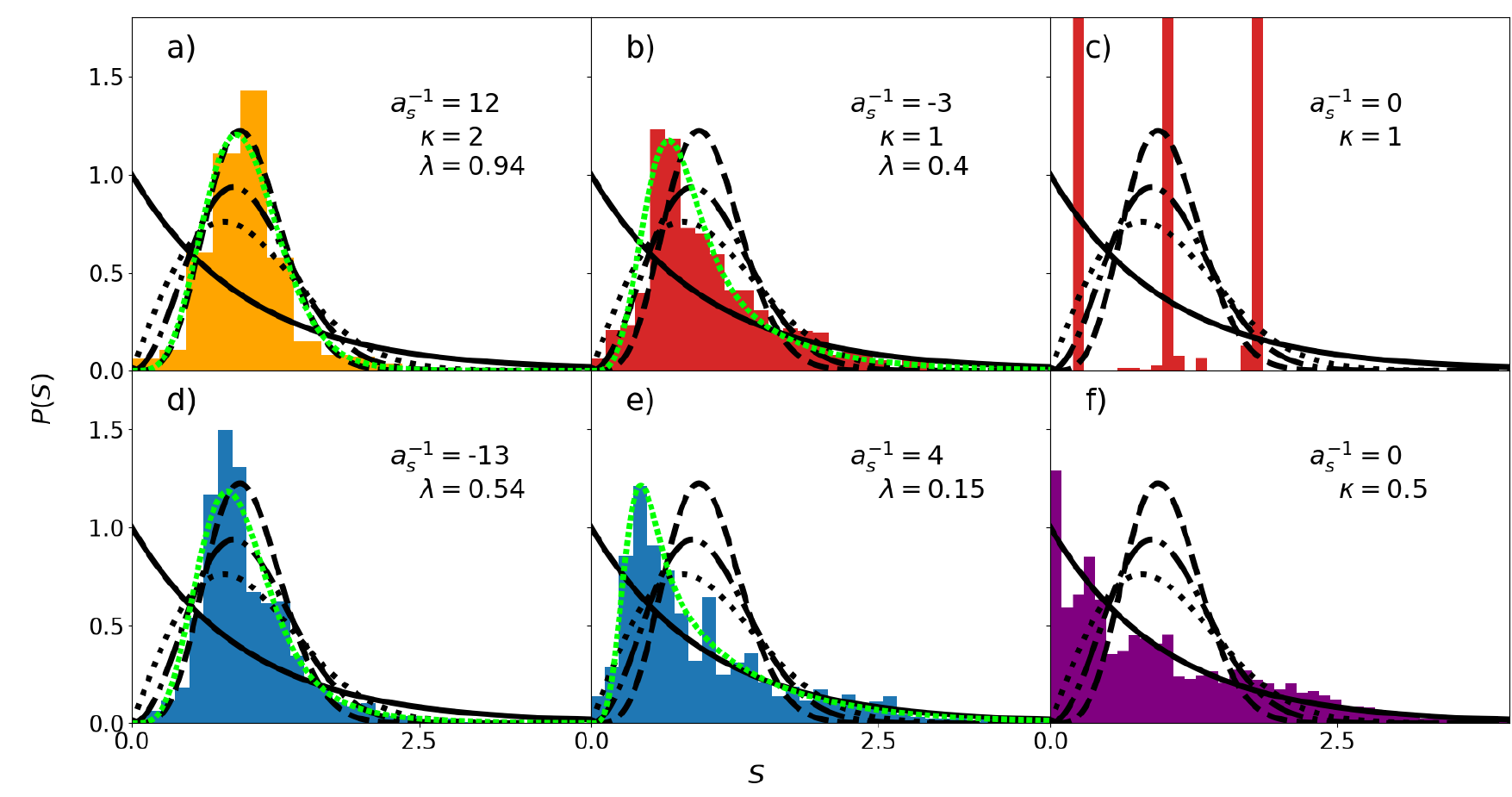}
    \caption{
    Transition from strong level repulsion to integrable statistics in the quantum three-body problem. Shown are the level-spacing distributions $P(S)$ for two indistinguishable fermions and impurity in panels (a) for light ($\kappa=2$, orange), (b) and (c) for equal ($\kappa=1$, red), and (f) for heavy ($\kappa=0.5$, purple) impurity mass. Panels (d) and (e) are for three indistinguishable bosons (blue). Scattering lengths $a_s$ are as indicated.   
    Lines show the predictions by RMT for integrable (Poissonian, full black line), quantum chaotic GSE (dashed black line), GUE (dot-dashed black line), and GOE (sparsely dotted black line). The predictions of the Rosenzweig-Porter for a best-fit $\lambda$ are given by the densely dotted green line. 
    Histograms in (c) and (f) are comprised of $\approx19000$ levels and in (a) of $\approx360$;
    the rest are comprised of $\approx1100$.
    It is more difficult to accurately obtain large numbers of energy levels away from unitarity, particularly for small positive $a_s$. All data correspond to zero angular momentum. $\kappa=m_{\rm fermion}/m_{\rm impurity}$, $\kappa=1$ for bosons. }
    \label{fig:P(S)Grid}
\end{figure*}

Now we are ready to examine the level-spacing statistics of the three-body system. 
In Fig.~\ref{fig:P(S)Grid} we plot the level-spacing distributions for various system parameters for zero angular momentum. For finite $a_s$ the behavior is qualitatively similar for bosons and fermions with any impurity mass. The left-hand panels of Fig.~\ref{fig:P(S)Grid} correspond to small $a_{s}$, far from unitarity, and show level-spacing distributions consistent with the predictions of the Gaussian Symplectic Ensemble (GSE, $\beta=4$) of RMT~\cite{Mehta1990}. The Gaussian Orthogonal (GOE, $\beta=1)$ and Unitary (GUE, $\beta=2$) Ensemble predictions are shown for comparison but fail to capture the strong suppression of the histograms at low $S$. The middle panels correspond to moderate $|a_s|$. Here the distributions' peaks have moved closer to zero and the tails have lengthened, approaching a Poissonian distribution.


These intermediate distributions are well described by the Rosenzweig-Porter model~\cite{rosenzweig1960repulsion, vcadevz2024rosenzweig}. In this model, the Hamiltonian has regular and chaotic contributions, $H=H_{\rm reg}+\lambda \Gamma_{N}  H_{\rm chaos}$, with $\lambda$ parameterizing the relative size of those contributions and $\Gamma_N$ ensures the unfolded spectrum is independent of the Hilbert space dimension $N$. The level-spacing distribution for a given $\lambda$, $P_{\lambda}(S)$, is known analytically~\cite{vcadevz2024rosenzweig,Schierenberg2012}. In the left and middle panels of Fig.~\ref{fig:P(S)Grid} we plot the fitted $P_{\lambda}(S)$ on top of the finite $a_s$ level-spacing distributions. In the left hand panels the small $a_s$ level-spacing distribution and the fitted $P_{\lambda}(S)$ are both close to the GSE prediction. In the middle panels the intermediate $a_s$ distributions don't match the GSE prediction but are well described by the fitted $P_{\lambda}(S)$ distributions. The right hand panels of Fig.~\ref{fig:P(S)Grid} show the unitary regime ($a_s\rightarrow\infty$), where the distributions are Poissonian or a ``stick distribution'', elaborated upon below.

\begin{figure}[h]
    \centering
    \includegraphics[width=8.5cm, height=5cm]{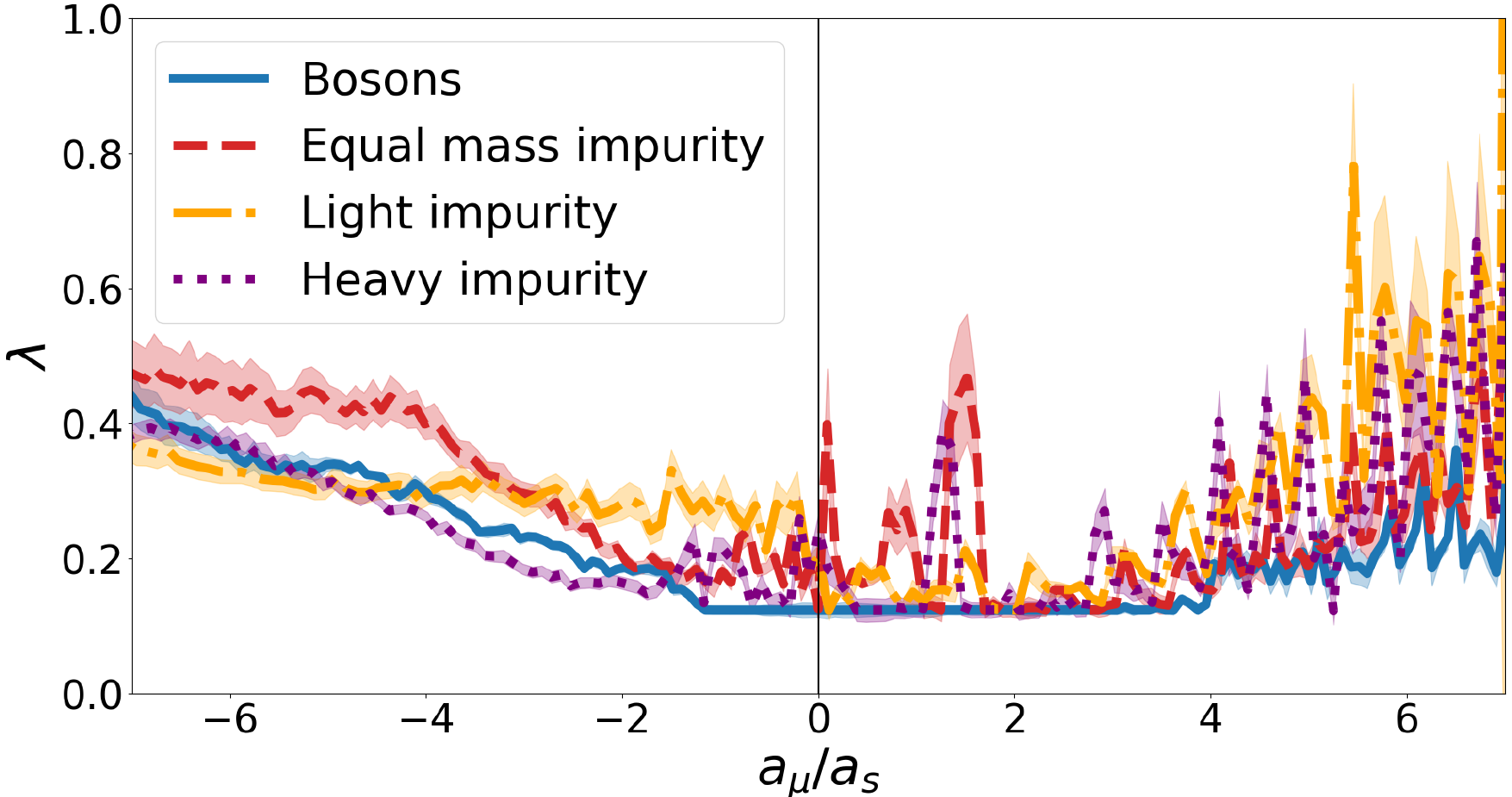}
    \caption{Best-fit value for the Rosenzweig-Porter parameter $\lambda$ as a function of $a_s$. The solid blue line corresponds to bosons, the dashed red line to fermions with $\kappa=1$, the dot-dashed orange line to $\kappa=2$, and the dotted purple to $\kappa=0.5$. The shading indicates one standard deviation in the $\lambda$ fit. All data correspond to zero angular momentum.}
    \label{fig:LambdaVsAs}
\end{figure}

By finding the best-fit $\lambda$ for different $a_s$ we can quantify the relationship between interaction strength and the degree of chaos. In Fig.~\ref{fig:LambdaVsAs} we plot the fitted value of $\lambda$ against $a_{s}$. The behavior is similar across particle symmetry and impurity masses. Namely, $\lambda$ is minimized near unitarity and increases as the interactions weaken. The increase in $\lambda$ with $a_s^{-1}>0$ is notably less smooth. This is because the energy spectrum varies more quickly for positive $a_s$. Note, for $\lambda\lesssim0.1$ the fitting procedure becomes less reliable due to a singularity at $\lambda=0$ in $P_{\lambda}(S)$,~\cite{vcadevz2024rosenzweig} meaning $\lambda$ is overstated close to unitarity. 

So far we have neglected the unitary case. At unitarity there is a subtlety regarding symmetry. Typically, one considers a single symmetry subspace in isolation. We have done this above by neglecting the center-of-mass motion and focusing on the zero angular momentum ($l=0$) spectrum. However, in the unitary and non-interacting limits there is an SO(2,1) symmetry~\cite{Werner2006unitarygas}. Accounting for this divides the unitary spectrum into subspectra which are evenly spaced ladders of states. However, it is still appropriate to consider all these subspectra together (the undivided unitary spectrum) as a limit of the finite $a_s$ spectrum. 

In Fig.~\ref{fig:P(S)Grid} (f) the level-spacing distribution for fermions with a heavy impurity ($\kappa\equiv m_{\rm fermion}/m_{\rm impurity}=0.5$) is a Poissonian, indicating regular dynamics. However for $\kappa=1$ in panel (c) the distribution is clearly not a Possonian. A few discrete spacings are favored. Such a distribution (sometimes termed ``stick statistic'') is unexpected for a typical regular quantum system but still indicates regularity. It is reminiscent of the non-interacting simple harmonic oscillator distribution which only allows a single level spacing~\cite{Berry1977, Drod1991, Chakrabarti2003}. Whether or not the unitary level spacing distribution is a Poissonian depends the mass ratio. Specifically, it depends on whether or not
\begin{eqnarray}
K=\arctan\left(\kappa^{-1}\sqrt{1+2\kappa}\right)
\end{eqnarray}
is commensurate with $\pi$. For example, stick statistics results from $\kappa=1$ or $\kappa=1+\sqrt{2}$ ($K=\pi/3$ or $K=\pi/4$) and Poisson statistics results from e.g. $\kappa=2$ ($K=0.841\dots$). This result derives from the transcendental equation [Eq.~\eqref{eq:Transcendental}, see the appendix] that determines the unitary energies. The unitary energies are $E=(2q+\alpha_{n,l}+1)\hbar\omega$ where $q\in\mathbb{Z}_{\geq0}$ and $\alpha_{n,l}$ is determined by Eq.~\eqref{eq:Transcendental}. It can be shown that for large $n$
\begin{align} \label{eq:AlphaApproach}
\nonumber
    \alpha_{n,l} {\, \rm mod \, 2} \to
    &\; \frac{\eta(-1)^l(1+\kappa)^2}{\kappa\sqrt{1+2\kappa}}\frac{\cos[K(2n+\phi)]}{2n+\phi} +\delta_{l, \rm even}.\\
\end{align}
Where $\phi$ is a constant phase offset and $\eta=-1(2)$ for fermions(bosons).

Energy levels with the same $\alpha_{n,l}$ are separated by $2\hbar\omega$, meaning each state within a cluster corresponds to a different $\alpha_{n,l}$.  Equation \eqref{eq:AlphaApproach} then controls the relative positions of the states within each cluster. If the ratio $K/\pi$ is a rational number,
then the cosine part of Eq.~\eqref{eq:AlphaApproach} is periodic with integer $n$ and only a few distinct level spacings are allowed (after unfolding). Hence the discretized distribution. In the incommensurate case the cosine part is not periodic with integer $n$, and any level spacing is allowed. This is the origin of the stick and Poisson statistics at unitarity. However, Eq.~\eqref{eq:AlphaApproach} only holds in the large $n$ limit. For small $n$ the true energies are different, resulting in a small number of off-peak level spacings.

\section{Long-range correlations }
\begin{figure}
    \centering
    \includegraphics[width=8cm, height=5cm]{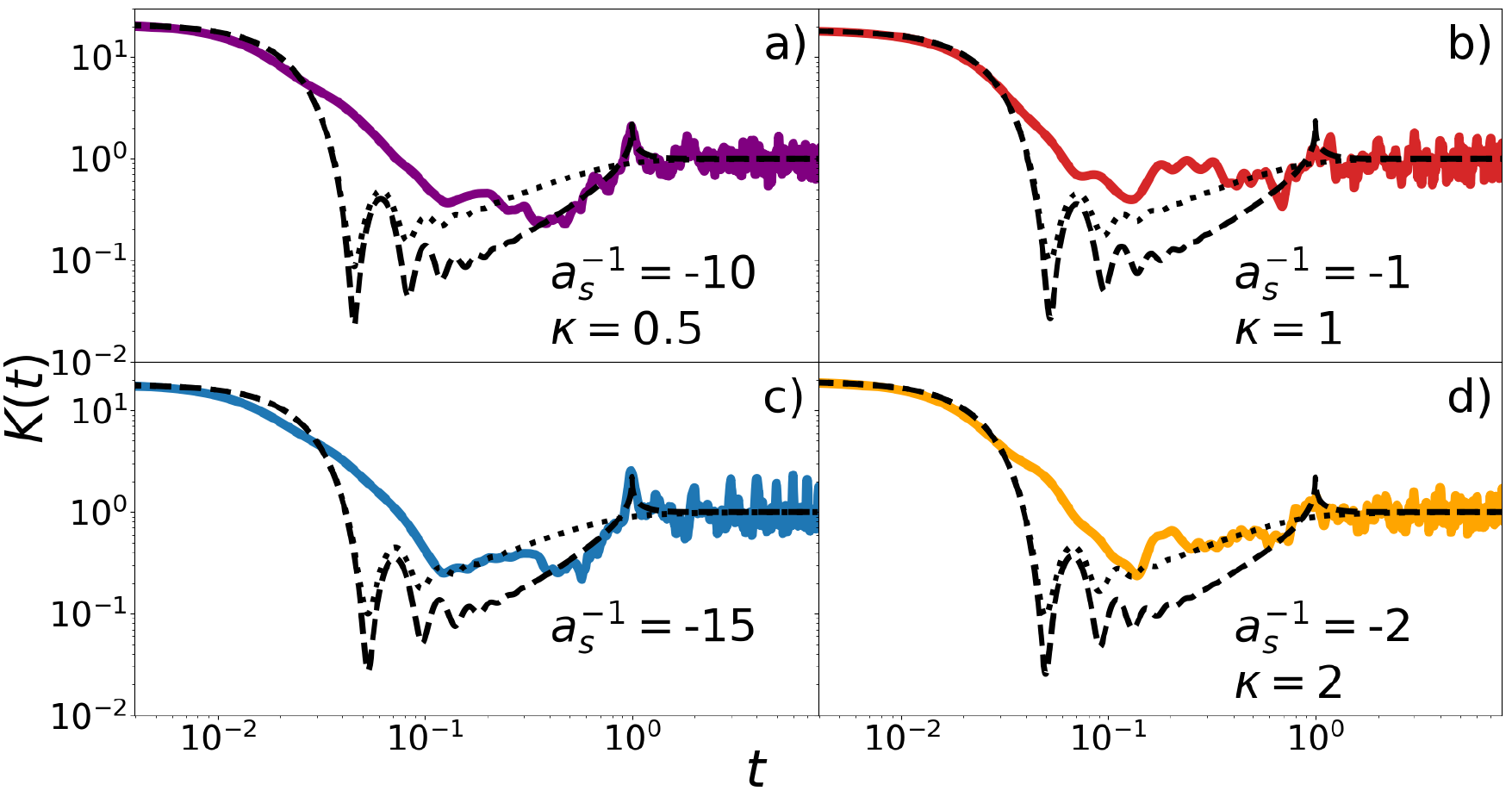}
    \caption{The spectral form factor $K(t)$ of Eq.~\eqref{eq:SFF} for various system parameters. The solid lines show data for two fermions with (a) heavy ($\kappa = 0.5$, purple), (b) equal mass ($\kappa=1$, red) and (d) light impurity ($\kappa = 2$, orange), and for bosons in panel (c) (blue).
    The dashed black lines show the GSE prediction and the dotted black lines the GOE prediction.}
    \label{fig:SFF}
\end{figure}

In addition to the short-range nearest-neighbor statistics of the spectrum discussed so far, RMT~\cite{Mehta1990}
also makes predictions about long-range correlations~\cite{brody1981random, guhr1998random, abul2014unfolding}. One important characteristic is the spectral form factor~\cite{French1988}. It is the Fourier transform of the two-point correlation function which encapsulates all two-point correlations in the system, not just the nearest neighbors~\cite{Mehta1990, liu2018spectral}.
It is given by
\begin{eqnarray}
   K(t)= Z^{-1}\left\langle \left\lvert N_{\rm max}^{-1}\small\sum_{n}^{N_{\rm max}}e^{2\pi i \epsilon_{n}t}\right\rvert^2 \right\rangle. \label{eq:SFF}
\end{eqnarray}
The normalization factor $Z$ ensures $K(t\gg1)\approx 1$. For systems that depend on some parameter, the average, $\langle \cdot \rangle$, is usually taken over an ensemble of different realizations of the energy spectrum~\cite{fogarty2021probing, vsuntajs2020quantum, li2024spectral}. Here, we average over different clusters of states. We neglect clusters with fewer than five levels. The total number of clusters included in the calculation for Fig.~\ref{fig:SFF} is approximately 45 depending on $a_s$, corresponding to approximately 1000 levels in total.

In regular systems $K(t)$ approaches its long-time average from above~\cite{prakash2021universal}. For chaotic systems $K(t)$ quickly falls below its long-time average and ramps back up to it. This feature is called a correlation hole~\cite{Leviandier1956,torres2017dynamical, lydzba2022signatures, zarate2023generalized, das2025proposal}. In the GSE case a logarithmic singularity causes $K(t)$ to briefly exceed its long-time average at the end of the correlation hole. This feature clearly distinguishes it from the two other universality classes within Dyson's threefold way, the Gaussian orthogonal (GOE) and unitarity (GUE) ensembles~\cite{Dyson1962}.

In Fig.~\ref{fig:SFF} we plot the spectral form factor for various system parameters. Near unitarity, the right hand panels (b) and (d) of Fig.~\ref{fig:SFF}, the correlation hole is shallow or absent, and there is no apparent logarithmic singularity. Far from unitarity, the left-hand panels (a) and (c) of Fig.~\ref{fig:SFF}, the ramp out of the correlation hole matches well with the GSE prediction, and there is a peak matching the logarithmic singularity.

\begin{figure}[h]
\includegraphics[width=8cm, height=5cm]{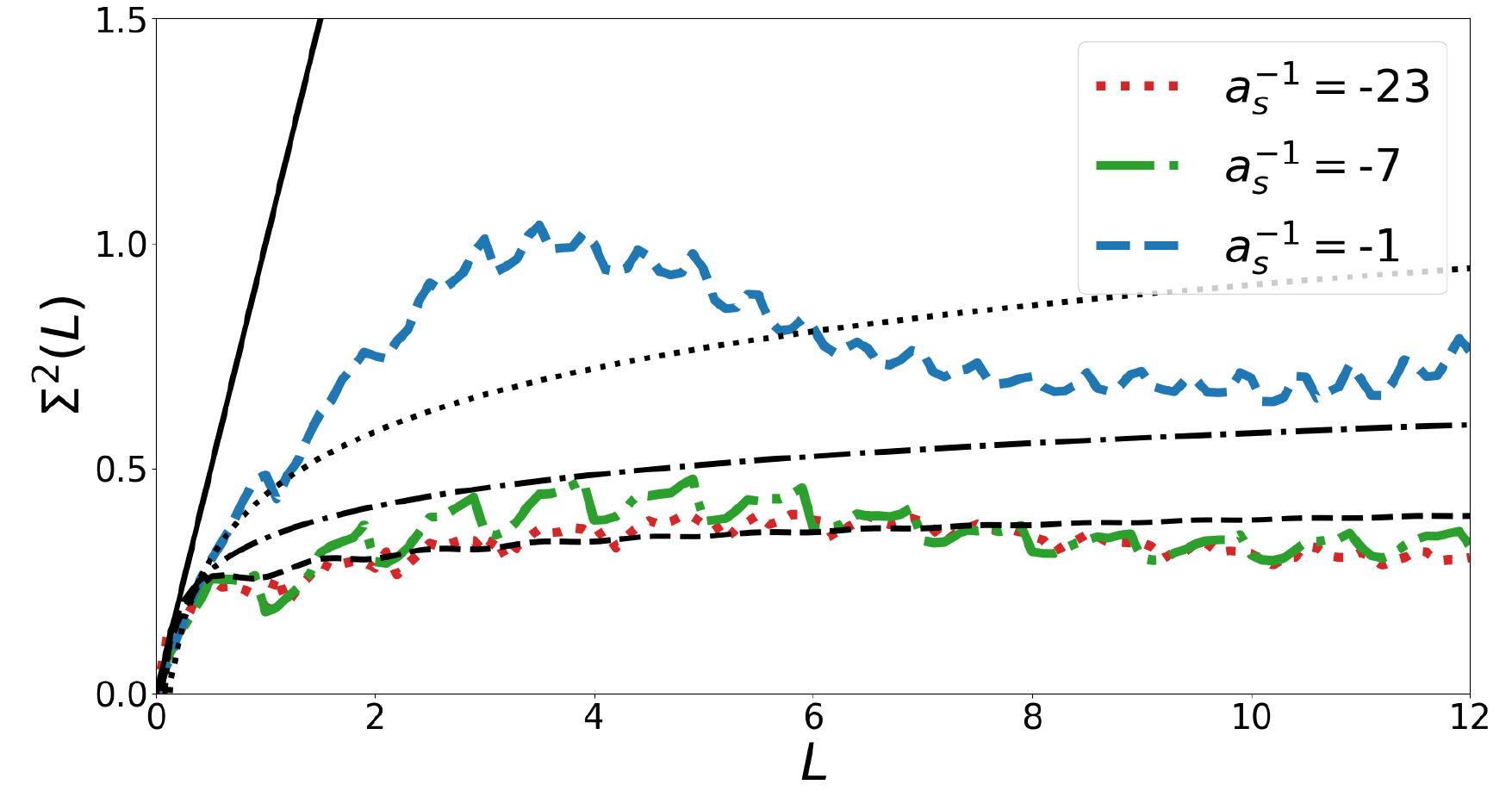}
\caption{The number variance, Eq.~\eqref{eq:NumVar}, for fermions with $\kappa=1$ and $l=0$ at various scattering lengths. The solid black straight line corresponds to a regular system, the dotted to GOE, the dot-dashed to GUE and the dashed to GSE.}
\label{fig:LongRange}
\end{figure}

Another characteristic of long-range correlations in the energy spectrum with definitive predictions from RMT
is the number variance \cite{Mehta1990, berry1988semiclassical, guhr1998random}
\begin{eqnarray}
\Sigma^{2}(L)=\langle N_E(L)^{2} \rangle-\langle N_E(L)\rangle^{2}\label{eq:NumVar},
\end{eqnarray}
where $N_{E}(L)$ is the number of unfolded levels in the range $[E,E+L]$. It characterizes the tendency for level bunching or repulsion at different scales $L$, and thus the ``rigidity'' of the spectrum. The average, $\langle \cdot \rangle$, is taken over $E$. For regular systems $\Sigma^{2}(L)$ is linear in $L$, and for chaotic ones it grows logarithmically with a decaying oscillatory component. In Fig.~\ref{fig:LongRange} we plot the number variance for a few illustrative examples. We again see the transition from Poisson to GSE statistics. 

\section{Discussion}
These results for the spectral properties of the interacting three-body system clearly indicate that it displays GSE statistics. This is unusual and unexpected, because GSE statistics are typically associated with time-reversal invariant spin-half systems~\cite{Haake2018, Mehta1990, Scharf1988} and has been observed experimentally for systems with strong spin-orbit coupling~\cite{Sacha2001,Kuemmeth2008}. Yet the present system is spinless and invariant under time-reversal operation. This is typically indicative of linear level repulsion, corresponding to the GOE. However, it is possible to have GSE statistics in spinless systems if it exhibits symplectic symmetry, i.e., is quaternion real,~\cite{joyner2014gse, rehemanjiang2016microwave, Lu2020, che2025experimental} 
implying that every energy level is doubly degenerate (Kramers degeneracy), which is not observed in this work. These systems are realized with quantum graphs~\cite{Kottos1999} that are constructed such that they have the same structure as a Hermitean Hamiltonian with symplectic symmetry consisting of two diagonal blocks containing a GUE matrix and its complex conjugate, respectively, and the off-diagonal blocks corresponding to an complex, antisymmetric matrix. We were not able to identify such a structure in the Hamiltonian analyzed in this work. Examples for time-reversal invariant quantum systems with chaotic classical dynamics, that do not comply with BGS are systems with a discrete rotational symmetry~\cite{Robbins1989,keating1997discrete} or with a unidirectional classical dynamics like, e.g., constant widths billiards~\cite{Knill1998,Gutkin2007,Dietz2014} and unidirectional quantum graphs~\cite{Akila2015,Che2022}. They comprise spectra that exhibit GUE instead of GOE statistics. Undirectional quantum graphs can also be designed such that their spectra comprise subspectra exhibiting GSE statistics with four-fold degeneracy~\cite{akila2019gse}.
The quantum three-body system of this Letter exhibits rotation and reflection symmetries due to the spherical trap and exchange symmetries due to identical fermions/bosons. Additionally, while the SO(2,1) symmetry only holds in the unitary and non-interacting limits it may be only weakly broken for finite $a_s$. Yet it is difficult to identify which of these symmetries, or combination of them, is responsible for the GSE statistics. Alternatively, GSE statistics can be realized by ignoring every second eigenvalue of a GOE spectrum~\cite{Mehta1990}. This would imply that we missed every second eigenvalue. However, this appears to be untrue.

To summarize, we investigate the level-spacing statistics of three contact-interacting particles. We find strong numerical evidence of chaos for weak interactions. The level-spacing distribution interpolates between a Poissonian close to unitarity and a GSE distribution for weak interactions. This is well described by the Rosenzweig-Porter model. We further investigate long-range correlations via the number variance and spectral form factor, and similarly found a transition between Poisson and GSE statistics. These findings do not comply with the BGS conjecture, since the system neither belongs to the symplectic universality class nor is every second energy level missing. We suppose that the presence of GSE statistics results from the particular symmetry properties of the system, 
but additional numerical work is needed to find out which of them or which combination thereof is responsible.
Lastly, this system can be experimentally realized with cold neutral atoms (e.g.~in microtraps~\cite{serwane2011deterministic, zurn2012fermionization, zurn2013pairing, murmann2015two,reynolds2020direct}) where the short-range van-der-Waals interaction is well approximated by a contact interaction, or by realizing the dilute limit \cite{Werner2012} of an optical lattice quantum simulator \cite{Gross2017,Schafer2020}. The spectral from factor can be measured by way of the survival probability~\cite{torres2017dynamical, torres2018generic, zarate2023generalized, daug2023many}. However, this requires preparing the system in an even superposition of all states of the symmetry subspace in question. In the system considered in this work this means preparing the system in an even superposition of states of one particular center-of-mass eigenstate and one particular relative angular momentum quantum number. It is unclear how such a state could be prepared experimentally.


\begin{acknowledgments}
With thanks to Thomas Woodrow-Smith for providing part of the computer code used to calculate the non-unitary three-body energy spectrum. 
We thank Sergej Flach, Sandro Wimberger, Yvan Castin, Felix Werner, Maxim Ol'shanni and Vladimir Yurovski for interesting discussions.
This work was supported by the Marsden Fund of New
Zealand (contract no.\ MAU2007) from government funding
administered by the Royal Society Te Ap\=arangi. BD acknowledges financial support from the Institute for Basic Science (IBS) in the Republic of Korea through the Project IBS-R024-D1. This research would not have been possible without the computer cluster of the Centre for Theoretical Chemistry and Physics (CTCP) at Massey University.

\end{acknowledgments}

\appendix*
\renewcommand{\theequation}{A\arabic{equation}}
\section{}

We briefly review the derivations of the three-body relative energy spectrum. For finite $a_s$ we consider an ansatz wave function of the form~\cite{liu2010three}
\begin{eqnarray}
&&\psi_{\rm 3b}^{\rm rel}(\vec{r},\vec{\rho})=\nonumber\\
&&\qquad(1+\hat{Q})\sum_{n=0}^{\infty}C_{n}\psi_{\rm 2b}^{\rm rel}(\nu_{nl},r)N_{nl}R_{nl}(\rho)Y_{lm}(\hat{\rho}).\quad\label{eq:SummWavefunction}
\end{eqnarray}
Here, $\vec{r}=\vec{r}_2-\vec{r}_1$ and $\vec{\rho}=\gamma^{-1}[\vec{r}_3-(\vec{r}_1+\vec{r}_2)/2]$, where $\gamma=\sqrt{(1+2\kappa)/(1+\kappa)^2}$, describe the particles' positions in the relative frame. $C_n$ are coefficients of expansion, $\psi_{\rm 2b}^{\rm rel}(\nu_{nl},r)$ are the interacting two-body wave functions~\cite{busch1998two}, and $N_{nl}R_{nl}(\rho)Y_{lm}(\hat{\rho})$ are the non-interacting simple harmonic oscillator normalisation, radial, and angular wave functions respectively. $l$ is the angular momentum and $m$ is its projection. $\hat{Q}$ is the particle exchange operator which ensures the wave function has the appropriate bosonic/fermionic symmetry. For three identical bosons $\hat{Q}=\hat{P}_{13}+\hat{P}_{23}$, and for two identical fermions plus an impurity $\hat{Q}=-\hat{P}_{23}$, where particle one is the impurity and $\hat{P}_{ij}f(\vec{r}_{i},\vec{r}_{j},\vec{r}_{k})=f(\vec{r}_{j},\vec{r}_{i},\vec{r}_{k})$.

Combining Eqs.~\eqref{eq:BethePeierls}
 \&~\eqref{eq:SummWavefunction} leads to the matrix equation
\begin{eqnarray}
\frac{a_{\mu}}{a_{s}}
\begin{bmatrix}
C_{0}\\
C_{1}\\
\vdots
\end{bmatrix}
=
\begin{bmatrix}
X_{00l} & X_{01l} & X_{02l} & \dots\\
X_{10l} & X_{11l} & X_{12l} & \dots\\
\vdots & \vdots & \vdots & \ddots\\
\end{bmatrix}
\begin{bmatrix}
C_{0}\\
C_{1}\\
\vdots
\end{bmatrix}.
\label{eq:MatrixEquation}
\end{eqnarray}
Where
\begin{eqnarray}
X_{n'nl}&=&
\frac{2\Gamma(-\nu_{n'l})}{\Gamma(-\nu_{n'l}-\frac{1}{2})}\delta_{n'n}
-\eta\frac{(-1)^l}{\sqrt{\pi}}A_{n'nl},\label{eq:MatrixEntries}\\
A_{n'nl}&=&a_{\mu}^3N_{n'l}N_{nl} \left(\frac{\kappa}{1+\kappa}\right)^l\nonumber\\
& \times& \int_{n}^{\infty}\tilde{\rho}^{2+2l}e^{-\tilde{\rho}^2}L_{n'}^{l+1/2}(\tilde{\rho}^2)L_{n}^{l+1/2}\left(\frac{\kappa^2 \tilde{\rho}^2}{(1+\kappa)^2}\right)\nonumber\\
&&\Gamma(-\nu_{nl})U\left(-\nu_{nl},\frac{3}{2},\frac{1+2\kappa}{(1+\kappa)^2}\tilde{\rho}^2 \right)d\tilde{\rho},
\end{eqnarray}
$a_{\mu}=\sqrt{\hbar/\mu\omega}$ is the simple harmonic oscillator length-scale, and $\mu=mm_{i}/(m+m_{i})$. The $\nu_{n'l}$s are determined by an arbitrarily chosen energy, $E_{\rm rel}=(2\nu_{n'l}+2n+l+3)\hbar\omega$, and Eq.~\eqref{eq:MatrixEquation} determines the scattering lengths for which the chosen energy is an eigenenergy as an eigenvalue problem. In this way the whole spectrum can be constructed.

In the unitary limit ($a_{s}^{-1}\rightarrow 0$) and the non-interacting case ($a_s=0$), all of the Hamiltonian's eigenstates can be described with exact closed form solutions~\cite{werner2006unitary, liu2010three}, implying integrability. The relative wavefunctions can be written
\begin{eqnarray}
    \psi_{3\rm b}^{\rm rel}(R,\theta,\vec{\Omega})\propto\frac{F_{q\alpha_{n,l}}(R)}{R^2}(1+\hat{Q})\frac{\varphi_{l\alpha_{n,l}}(\theta)}{\sin(2\theta)}Y_{lm}(\hat{\rho}),\qquad\label{eq:HypergeoWaveFunc}
\end{eqnarray}
where $2R^2=r^2+\rho^2$ and $\theta=\arctan(r/\rho)$. $F_{q\alpha_{n,l}}$ is a product of a Gaussian and a polynomial or Whittaker function depending on $\alpha_{n,l}$, and $\varphi_{l\alpha_{n,l}}$ is a product of trigonometric and hypergeometric functions. The energies are given by
\begin{eqnarray}
E_{\rm rel}= (2q+\alpha_{n,l}+1)\hbar\omega,\label{eq:UnivUnitEn}
\end{eqnarray}
where $q=0,1,2\dots$ and $\alpha_{n,l}$ are the solutions to the transcendental equation derived from applying the Bethe-Peierls boundary condition to Eq.~\eqref{eq:HypergeoWaveFunc}. 
\begin{eqnarray}
&&\frac{-\rho\varphi(0)Y_{lm}(\hat{\rho})}{2a_{s}}=\frac{d\varphi_{l,\alpha_{n,l}}}{d\theta}\Big |_{\theta=0}\nonumber\\
&&+\eta(-1)^l\dfrac{(1+\kappa)^2}{\kappa\sqrt{1+2\kappa}}\varphi_{l,\alpha_{n,l}}\left[\arctan\left(\frac{\sqrt{1+2\kappa}}{\kappa}\right)\right],\nonumber\\\label{eq:Transcendental}
\end{eqnarray}

Equation~\eqref{eq:Transcendental} only well defines $\alpha_{n,l}$ if the position dependence on the left-hand side drops out of the problem. i.e. if $a_{s}$ is zero (non-interacting) or infinitely large (unitarity). For finite $a_s$ this approach cannot produce exact closed form descriptions of the eigenstates. Some illustrative values of $\alpha_{n,0}$ in the unitary limit are presented in Table \ref{tab:SVals}.

%
%
%
%
%
%

\begin{table}[b]
\caption{Values of $\alpha_{n,0}$ for bosons and mass balanced fermions.}
\label{tab:SVals}
\begin{ruledtabular}
\begin{tabular}{ccc}
\textrm{n} &
\textrm{bosons}&
\textrm{fermions $\kappa=1$}\\
\colrule
0 & i$\cdot$ 1.006\dots & 2.166\dots\\
1 & 4.465\dots & 5.127\dots\\
2 & 6.818\dots & 7.114\dots\\
\vdots & \vdots & \vdots\\
99 & 201.0146\dots & 200.9926 \dots \\
100 & 202.9927\dots & 203.0036 \dots\\
\end{tabular}
\end{ruledtabular}
\end{table}

For equal-mass fermions, the solutions of Eq.~\eqref{eq:Transcendental} for $\alpha_{n,l}$ are all real and exhaust the spectrum of the Hamiltonian defined by Eqs.~\eqref{eq:Hamiltonian} \& \eqref{eq:BethePeierls}. For bosons, and fermions with some some combinations of $l$, $\kappa$, and $\eta$ \cite{kerin2023energetics}, however, imaginary solutions exist (e.g. $\alpha_{0,0}=i\cdot 1.006$\dots for  bosons). Note the $\alpha_{n,l}$s are not eigenvalues of the Hamiltonian, but $\alpha_{n,l}^2-4$ are. These imaginary values of $\alpha_{n,l}$ indicate non-universal Efimov-like states~\cite{efimov1971bound}, whose energies do not obey Eq.~\eqref{eq:UnivUnitEn} and which are not uniquely determined by the two-body Bethe Peierls boundary condition \eqref{eq:BethePeierls} \cite{Werner2006unitarygas,Blume2018,kerin2023energetics}. An additional boundary condition, or three-body parameter $R_{t}$, has to be introduced to make the Hamiltonian well defined. The energy is then given by
\begin{eqnarray}
-|\alpha_{n,l}|\ln\left(\frac{R_{t}}{a_{\mu}}\right)&=&\arg\left[ \frac{\Gamma\left(\dfrac{1+\alpha_{n,l}-E_{\rm rel}}{2}\right)}{\Gamma(1+\alpha_{n,l})} \right] \quad {\rm mod}\;\pi .\nonumber\\
\label{eq:EfimovEnergy}
\end{eqnarray}
At higher energies, the significant minority of states are Efimov states. The number of states with energy less than $E$ grows quadratically with $E$ for non-Efimov states and linearly for Efimov states. 

The variational ansatz of Eq.~\eqref{eq:SummWavefunction} used for finite scattering length can capture both the universal states and the non-universal Efimov states, where the precise energies of the Efimov states is fixed by the finite dimension of the matrix Eq.~\eqref{eq:MatrixEquation} (providing an effective three-body parameter)~\cite{kerin2023energetics}.
For the spectral statistics reported in this work, the Efimov states appear to play no major role, as their number is much smaller than those of the universal states. This is further supported by the fact that the $\beta=4$ statistical signatures are observed in both cases where Efimov states are present (bosons) and where they are not (fermions).

\bibliography{Chaos-Refs,manual_refs}

@article{Blume2018,
  title = {Harmonically Trapped Four-Boson System},
  author = {Blume, D. and Sze, M. W. C. and Bohn, J. L.},
  year = {2018},
  month = mar,
  journal = {Physical Review A},
  volume = {97},
  number = {3},
  pages = {033621},
  publisher = {American Physical Society},
  doi = {10.1103/PhysRevA.97.033621},
  urldate = {2025-10-10}
}

@Article{Bohigas1984,
  Title                    = {Characterization of Chaotic Quantum Spectra and Universality of Level Fluctuation Laws},
  Author                   = {Bohigas, O. and Giannoni, M. J. and Schmit, C.},
  Journal                  = {Phys. Rev. Lett.},
  Year                     = {1984},

  Month                    = {Jan},
  Number                   = {1},
  Pages                    = {1},
  Volume                   = {52},
  Doi                      = {10.1103/PhysRevLett.52.1},
  Publisher                = {American Physical Society}
}

@article{Scharf1988,
        doi = {10.1209/0295-5075/5/5/001},
        year = 1988,
        month = {mar},
        publisher = {{IOP} Publishing},
        volume = {5},
        number = {5},      
        pages = {383--389},
        author = {R Scharf and B Dietz and M Ku{\'s} and F Haake and M. V Berry},
        title = {Kramers{\textquotesingle} Degeneracy and Quartic Level Repulsion},
        journal = {Europhys. Lett. ({EPL})}

}

@article{Sacha2001,
  title = {Driven Rydberg Atoms Reveal Quartic Level Repulsion},
  author = {Sacha, Krzysztof and Zakrzewski, Jakub},
  journal = {Phys. Rev. Lett.},
  volume = {86},
  issue = {11},
  pages = {2269--2272},
  numpages = {0},
  year = {2001},
  month = {Mar},
  publisher = {American Physical Society},
  doi = {10.1103/PhysRevLett.86.2269},
  url = {https://link.aps.org/doi/10.1103/PhysRevLett.86.2269}
}

@article{Kuemmeth2008,
author = {Kuemmeth, Ferdinand and Bolotin, Kirill I. and Shi, Su-Fei and Ralph, Daniel C.},
title = {Measurement of Discrete Energy-Level Spectra in Individual Chemically Synthesized Gold Nanoparticles},
journal = {Nano Lett.},
volume = {8},
number = {12},
pages = {4506-4512},
year = {2008},
doi = {10.1021/nl802473n},
}

@article{Guhr1989,
title = {Coexistence of collectivity and chaos in nuclei},
journal = {Ann. Phys.},
volume = {193},
number = {2},
pages = {472 - 489},
year = {1989},
doi = {https://doi.org/10.1016/0003-4916(89)90006-7},
author = {T. Guhr and H. A. Weidenm\"uller},
}

@article{Weidenmueller2009,
  title = {Random matrices and chaos in nuclear physics: Nuclear structure},
  author = {Weidenm\"uller, H. A. and Mitchell, G. E.},
  journal = {Rev. Mod. Phys.},
  volume = {81},
  issue = {2},
  pages = {539--589},
  year = {2009},
  month = {May},
  publisher = {American Physical Society},
  doi = {10.1103/RevModPhys.81.539},
}

@Article{Zelev1996,
  Title                    = {Quantum chaos and complexity in nuclei},
  Author                   = {Vladimir Zelevinsky}, 
  Journal                  = {Ann. Rev. Nucl. Part. Sci.},
  Year                     = {1996},
  Pages                    = {237},
  Volume                   = {46},
  Doi                      = {10.1146/annurev.nucl.46.1.237},
}

@article{Kravtsov2015,
doi = {10.1088/1367-2630/17/12/122002},
year = {2015},
publisher = {IOP Publishing},
volume = {17},
number = {12},
pages = {122002},
author = {V E Kravtsov and I M Khaymovich and E Cuevas and M Amini},
title = {A random matrix model with localization and ergodic transitions},
journal = {New J. Phys.}

}

@Book{Haake2018,
  Title                    = {Quantum Signatures of Chaos},
  Author                   = {Haake, F. and Gnutzmann, S. and Ku\'s, Marek},
  Publisher                = {Springer-Verlag},
  Year                     = {2018},
  Address                  = {Heidelberg},
}

@article{Pal2010,
  title = {Many-body localization phase transition},
  author = {Pal, Arijeet and Huse, David A.},
  journal = {Phys. Rev. B},
  volume = {82},
  issue = {17},
  pages = {174411},
  numpages = {7},
  year = {2010},
  month = {Nov},
  publisher = {American Physical Society},
  doi = {10.1103/PhysRevB.82.174411},
  url = {https://link.aps.org/doi/10.1103/PhysRevB.82.174411}
}

@article{Serbyn2013,
  title = {Local Conservation Laws and the Structure of the Many-Body Localized States},
  author = {Serbyn, Maksym and Papi\ifmmode \acute{c}\else \'{c}\fi{}, Z. and Abanin, Dmitry A.},
  journal = {Phys. Rev. Lett.},
  volume = {111},
  issue = {12},
  pages = {127201},
  numpages = {5},
  year = {2013},
  month = {Sep},
  publisher = {American Physical Society},
  doi = {10.1103/PhysRevLett.111.127201},
  url = {https://link.aps.org/doi/10.1103/PhysRevLett.111.127201}
}

@article{Pino2019,
doi = {10.1088/1751-8121/ab4b76},
url = {https://dx.doi.org/10.1088/1751-8121/ab4b76},
year = {2019},
month = {oct},
publisher = {IOP Publishing},
volume = {52},
number = {47},
pages = {475101},
author = {Pino, M and Tabanera, J and Serna, P},
title = {From ergodic to non-ergodic chaos in Rosenzweig–Porter model},
journal = {Journal of Physics A: Mathematical and Theoretical}
}

@article{Drod1991,
  title = {Near-ground-state spectral fluctuations in multidimensional separable systems},
  author = {Dro\.zd\.z, S. and Speth, J.},
  journal = {Phys. Rev. Lett.},
  volume = {67},
  issue = {5}, 
  pages = {529--532},
  numpages = {0},
  year = {1991},
  month = {Jul},
  publisher = {American Physical Society},
  doi = {10.1103/PhysRevLett.67.529},
}

@Book{Mehta1990,  
  Title                    = {Random Matrices},
  Author                   = {Madan Lal Mehta},
  Publisher                = {Academic Press London},
  Year                     = {1990}
}

@article{Schierenberg2012, 
  title = {Wigner surmise for mixed symmetry classes in random matrix theory},
  author = {Schierenberg, Sebastian and Bruckmann, Falk and Wettig, Tilo},
  journal = {Phys. Rev. E},
  volume = {85},           
  issue = {6},
  pages = {061130},
  numpages = {24},
  year = {2012},
  month = {Jun},
  publisher = {American Physical Society},
  doi = {10.1103/PhysRevE.85.061130},
}

@article{Lu2020,
  title = {Experimental and numerical investigation of parametric spectral properties of quantum graphs with unitary or symplectic symmetry},
  author = {Lu, Junjie and Che, Jiongning and Zhang, Xiaodong and Dietz, Barbara},
  journal = {Phys. Rev. E},
  volume = {102},
  issue = {2},
  pages = {022309},
  numpages = {12},
  year = {2020},
  month = {Aug},
  publisher = {American Physical Society},
  doi = {10.1103/PhysRevE.102.022309},
}

@article{Dyson1962,
author = {Dyson,Freeman J. },
title = {The Threefold Way. Algebraic Structure of Symmetry Groups and Ensembles in Quantum Mechanics},
journal = {Journal of Mathematical Physics},
volume = {3},
number = {6},
pages = {1199-1215},
year = {1962},
doi = {10.1063/1.1703863},
}

@article{Deutsch1991,
  title = {Quantum statistical mechanics in a closed system},
  author = {Deutsch, J. M.},
  journal = {Phys. Rev. A},
  volume = {43},
  issue = {4},
  pages = {2046--2049},
  numpages = {0},
  year = {1991},
  month = {Feb},
  publisher = {American Physical Society},
  doi = {10.1103/PhysRevA.43.2046},
  url = {https://link.aps.org/doi/10.1103/PhysRevA.43.2046}
}

@article{Srednicki1994,
  title = {Chaos and quantum thermalization},
  author = {Srednicki, Mark},
  journal = {Phys. Rev. E},
  volume = {50},
  issue = {2},
  pages = {888--901},
  numpages = {0},
  year = {1994},
  month = {Aug},
  publisher = {American Physical Society},
  doi = {10.1103/PhysRevE.50.888},
  url = {https://link.aps.org/doi/10.1103/PhysRevE.50.888}
}

@article{Alessio2016,
author = {Luca D'Alessio and Yariv Kafri and Anatoli Polkovnikov and Marcos Rigol},
title = {From quantum chaos and eigenstate thermalization to statistical mechanics and thermodynamics},
journal = {Advances in Physics},
volume = {65},
number = {3},
pages = {239--362},
year = {2016},
publisher = {Taylor \& Francis},
doi = {10.1080/00018732.2016.1198134},
}

@phdthesis{werner2008trapped,
  title={Trapped cold atoms with resonant interactions: unitary gas and three-body problem},
  author={Werner, F{\'e}lix},
  school={Theses, Universit{\'e} Pierre et Marie Curieâ Paris VI, Paris France},
  year={2008}
}

@article{busch1998two,
  title={Two cold atoms in a harmonic trap},
  author={Busch, Thomas and Englert, Berthold Georg and Rza{\.z}ewski, Kazimierz and Wilkens, Martin},
  journal={Foundations of Physics},
  volume={28},
  number={4},
  pages={549--559},
  year={1998},
  publisher={Springer}
}

@article{werner2006unitary,
  title={Unitary quantum three-body problem in a harmonic trap},
  author={Werner, F{\'e}lix and Castin, Yvan},
  journal={Physical Review Letters},
  volume={97},
  number={15},
  pages={150401},
  year={2006},
  publisher={APS}
}

@article{Werner2006unitarygas,
  title = {Unitary gas in an isotropic harmonic trap: Symmetry properties and applications},
  author = {Werner, F\'elix and Castin, Yvan},
  journal = {Physical Review A},
  volume = {74},
  issue = {5},
  pages = {053604},
  numpages = {10},
  year = {2006},
  month = {Nov},
  publisher = {American Physical Society},
  doi = {10.1103/PhysRevA.74.053604},
  url = {https://link.aps.org/doi/10.1103/PhysRevA.74.053604}
}

@article{liu2009virial,
  title = {Virial Expansion for a Strongly Correlated Fermi Gas},
  author = {Liu, Xia Ji and Hu, Hui and Drummond, Peter D.},
  journal = {Physical Review Letters},
  volume = {102},
  issue = {16},
  pages = {160401},
  numpages = {4},
  year = {2009},
  month = {Apr},
  publisher = {American Physical Society},
  doi = {10.1103/PhysRevLett.102.160401},
  url = {https://link.aps.org/doi/10.1103/PhysRevLett.102.160401}
}

@article{liu2010three,
  title={Three attractively interacting fermions in a harmonic trap: Exact solution, ferromagnetism, and high-temperature thermodynamics},
  author={Liu, Xia Ji and Hu, Hui and Drummond, Peter D},
  journal={Physical Review A},
  volume={82},
  number={2},
  pages={023619},
  year={2010},
  publisher={APS}
}

@article{kestner2007level,
  title={Level crossing in the three-body problem for strongly interacting fermions in a harmonic trap},
  author={Kestner, J P and Duan, L M},
  journal={Physical Review A},
  volume={76},
  number={3},
  pages={033611},
  year={2007},
  publisher={APS}
}

@article{serwane2011deterministic,
  title={Deterministic preparation of a tunable few-fermion system},
  author={Serwane, Friedhelm and Z{\"u}rn, Gerhard and Lompe, Thomas and Ottenstein, TB and Wenz, AN and Jochim, S},
  journal={Science},
  volume={332},
  number={6027},
  pages={336--338},
  year={2011},
  publisher={American Association for the Advancement of Science}
}

@article{murmann2015two,
  title={Two fermions in a double well: Exploring a fundamental building block of the Hubbard model},
  author={Murmann, Simon and Bergschneider, Andrea and Klinkhamer, Vincent M and Z{\"u}rn, Gerhard and Lompe, Thomas and Jochim, Selim},
  journal={Physical Review Letters},
  volume={114},
  number={8},
  pages={080402},
  year={2015},
  publisher={APS}
}

@article{zurn2013pairing,
  title={Pairing in few-fermion systems with attractive interactions},
  author={Z{\"u}rn, G and Wenz, A. N. and Murmann, Simon and Bergschneider, Andrea and Lompe, Thomas and Jochim, S},
  journal={Physical Review Letters},
  volume={111},
  number={17},
  pages={175302},
  year={2013},
  publisher={APS}
}

@article{phillips1982laser,
  title={Laser deceleration of an atomic beam},
  author={Phillips, William D and Metcalf, Harold},
  journal={Physical Review Letters},
  volume={48},
  number={9},
  pages={596},
  year={1982},
  publisher={APS}
}

@article{chu1991laser,
  title={Laser manipulation of atoms and particles},
  author={Chu, Steven},
  journal={Science},
  volume={253},
  number={5022},
  pages={861--866},
  year={1991},
  publisher={American Association for the Advancement of Science}
}

@book{tannoudji1992atom,
  title={Atom-photon interactions},
  author={Tannoudji, Claude Cohen and Grynberg, Gilbert and Dupont-Roe, J},
  year={1992},
  publisher={New York, NY (United States); John Wiley and Sons Inc.}
}

@article{zurn2012fermionization,
  title={Fermionization of two distinguishable fermions},
  author={Z{\"u}rn, Gerhard and Serwane, Friedhelm and Lompe, T and Wenz, A. N. and Ries, Martin Gerhard and Bohn, Johanna Elise and Jochim, Selim},
  journal={Physical Review Letters},
  volume={108},
  number={7},
  pages={075303},
  year={2012},
  publisher={APS}
}

@article{bethe1935quantum,
  title={Quantum theory of the diplon},
  author={Bethe, Hans and Peierls, Rudolf},
  journal={Proceedings of the Royal Society of London. Series A-Mathematical and Physical Sciences},
  volume={148},
  number={863},
  pages={146--156},
  year={1935},
  publisher={The Royal Society London}
}

@article{efimov1971bound,
  title={Weakly-bound states of three resonantly-interacting particles,},
  author={Efimov, V.N},
  journal={Soviet Journal of Nuclear Physics},
  volume={12},
  pages={589-595},
  year={1971},
  publisher={}
}

@article{eismann2016universal,
  title={Universal loss dynamics in a unitary Bose gas},
  author={Eismann, Ulrich and Khaykovich, Lev and Laurent, S{\'e}bastien and Ferrier-Barbut, Igor and Rem, Benno S and Grier, Andrew T and Delehaye, Marion and Chevy, Fr{\'e}d{\'e}ric and Salomon, Christophe and Ha, Li-Chung and others},
  journal={Physical Review X},
  volume={6},
  number={2},
  pages={021025},
  year={2016},
  publisher={APS}
}

@article{kerin2023energetics,
  title={Energetics and Efimov states of three interacting bosons and mass-imbalanced fermions in a three-dimensional spherical harmonic trap},
  author={Kerin, A. D. and Martin, A. M.},
  journal={Journal of Physics B: Atomic, Molecular and Optical Physics},
  volume={56},
  number={5},
  pages={055201},
  year={2023},
  publisher={IOP Publishing}
}

@article{braaten2013universal,
  title={Universal relation for the inelastic two-body loss rate},
  author={Braaten, Eric and Hammer, HW},
  journal={Journal of Physics B: Atomic, Molecular and Optical Physics},
  volume={46},
  number={21},
  pages={215203},
  year={2013},
  publisher={IOP Publishing}
}

@article{torres2015dynamics,
  title={Dynamics at the many-body localization transition},
  author={Torres-Herrera, EJ and Santos, Lea F},
  journal={Physical Review B},
  volume={92},
  number={1},
  pages={014208},
  year={2015},
  publisher={APS}
}

@article{villasenor2024breakdown,
  title={Breakdown of the quantum distinction of regular and chaotic classical dynamics in dissipative systems},
  author={Villase{\~n}or, David and Santos, Lea F and Barberis-Blostein, Pablo},
  journal={Physical Review Letters},
  volume={133},
  number={24},
  pages={240404},
  year={2024},
  publisher={APS}
}

@mastersthesis{giasemis2022quantum,
  title={Quantum chaos in many-body systems without a classical analogue},
  author={Giasemis, Fotios Ioannis},
  year={2022},
  school={National Technical University of Athens}
}

@article{abul2014unfolding,
  title={Unfolding of the spectrum for chaotic and mixed systems},
  author={Abul-Magd, Ashraf A and Abul-Magd, Adel Y},
  journal={Physica A: Statistical Mechanics and its Applications},
  volume={396},
  pages={185--194},
  year={2014},
  publisher={Elsevier}
}

@article{abuelenin2018spectral,
  title={On the spectral unfolding of chaotic and mixed systems},
  author={Abuelenin, Sherif M},
  journal={Physica A: Statistical Mechanics and its Applications},
  volume={492},
  pages={564--570},
  year={2018},
  publisher={Elsevier}
}

@article{torres2017dynamical,
  title={Dynamical manifestations of quantum chaos: correlation hole and bulge},
  author={Torres-Herrera, E J and Santos, Lea F},
  journal={Philosophical Transactions of the Royal Society A: Mathematical, Physical and Engineering Sciences},
  volume={375},
  number={2108},
  pages={20160434},
  year={2017},
  publisher={The Royal Society Publishing}
}

@article{torres2018generic,
  title={Generic dynamical features of quenched interacting quantum systems: Survival probability, density imbalance, and out-of-time-ordered correlator},
  author={Torres-Herrera, E J and Garc{\'\i}a-Garc{\'\i}a, Antonio M and Santos, Lea F},
  journal={Physical Review B},
  volume={97},
  number={6},
  pages={060303},
  year={2018},
  publisher={APS}
}

@article{torres2017extended,
  title={Extended nonergodic states in disordered many-body quantum systems},
  author={Torres-Herrera, E Jonathan and Santos, Lea F},
  journal={Annalen der Physik},
  volume={529},
  number={7},
  pages={1600284},
  year={2017},
  publisher={Wiley Online Library}
}

@article{schiulaz2019thouless,
  title={Thouless and relaxation time scales in many-body quantum systems},
  author={Schiulaz, Mauro and Torres-Herrera, E Jonathan and Santos, Lea F},
  journal={Physical Review B},
  volume={99},
  number={17},
  pages={174313},
  year={2019},
  publisher={APS}
}

@article{de2020quantum,
  title={Quantum chaos in a system with high degree of symmetries},
  author={de la Cruz, Javier and Lerma-Hern{\'a}ndez, Sergio and Hirsch, Jorge G},
  journal={Physical Review E},
  volume={102},
  number={3},
  pages={032208},
  year={2020},
  publisher={APS}
}

@article{huber2021morphology,
  title={Morphology of three-body quantum states from machine learning},
  author={Huber, David and Marchukov, Oleksandr V and Hammer, Hans Werner and Volosniev, Artem G},
  journal={New Journal of Physics},
  volume={23},
  number={6},
  pages={065009},
  year={2021},
  publisher={IOP Publishing}
}

@Article{AnhTrai2023QuantumChaos,
  title={{Quantum chaos in interacting Bose-Bose mixtures}},
  author={Tran Duong Anh-Tai and Mathias Mikkelsen and Thomas Busch and Thomás Fogarty},
  journal={SciPost Phys.},
  volume={15},
  pages={048},
  year={2023},
  publisher={SciPost},
  doi={10.21468/SciPostPhys.15.2.048},
  url={https://scipost.org/10.21468/SciPostPhys.15.2.048},
}

@article{lydzba2022signatures,
  title={Signatures of quantum chaos in low-energy mixtures of few fermions},
  author={{\L}yd{\.z}ba, Patrycja and Sowi{\'n}ski, Tomasz},
  journal={Physical Review A},
  volume={106},
  number={1},
  pages={013301},
  year={2022},
  publisher={APS}
}

@article{reynolds2020direct,
  title={Direct measurements of collisional dynamics in cold atom triads},
  author={Reynolds, LA and Schwartz, E and Ebling, U and Weyland, M and Brand, J and Andersen, MF},
  journal={Physical Review Letters},
  volume={124},
  number={7},
  pages={073401},
  year={2020},
  publisher={APS}
}

@incollection{santos2018nonequilibrium,
  title={Nonequilibrium quantum dynamics of many-body systems},
  author={Santos, Lea F and Torres-Herrera, E Jonathan},
  booktitle={Chaotic, Fractional, and Complex Dynamics: New Insights and Perspectives},
  pages={231--260},
  year={2017},
  publisher={Springer}
}

@article{fogarty2021probing,
  title={Probing the edge between integrability and quantum chaos in interacting few-atom systems},
  author={Fogarty, Thom{\'a}s and Garc{\'\i}a-March, Miguel {\'A}ngel and Santos, Lea F and Harshman, Nathan L},
  journal={Quantum},
  volume={5},
  pages={486},
  year={2021},
  publisher={Verein zur F{\"o}rderung des Open Access Publizierens in den Quantenwissenschaften}
}

@article{de2024thermalization,
  title={Thermalization in Trapped Bosonic Systems With Disorder},
  author={de la Cruz, Javier and Diaz-Mejia, Carlos and Lerma-Hernandez, Sergio and Hirsch, Jorge G},
  journal={arXiv preprint arXiv:2407.04818},
  year={2024}
}

@article{alet2018many,
  title={Many-body localization: An introduction and selected topics},
  author={Alet, Fabien and Laflorencie, Nicolas},
  journal={Comptes Rendus Physique},
  volume={19},
  number={6},
  pages={498--525},
  year={2018},
  publisher={Elsevier}
}

@article{vsuntajs2020quantum,
  title={Quantum chaos challenges many-body localization},
  author={{\v{S}}untajs, Jan and Bon{\v{c}}a, Janez and Prosen, Toma{\v{z}} and Vidmar, Lev},
  journal={Physical Review E},
  volume={102},
  number={6},
  pages={062144},
  year={2020},
  publisher={APS}
}

@article{bulchandani2022onset,
  title={Onset of many-body quantum chaos due to breaking integrability},
  author={Bulchandani, Vir B and Huse, David A and Gopalakrishnan, Sarang},
  journal={Physical Review B},
  volume={105},
  number={21},
  pages={214308},
  year={2022},
  publisher={APS}
}

@article{santhanam2022quantum,
  title={Quantum kicked rotor and its variants: Chaos, localization and beyond},
  author={Santhanam, MS and Paul, Sanku and Kannan, J Bharathi},
  journal={Physics Reports},
  volume={956},
  pages={1--87},
  year={2022},
  publisher={Elsevier}
}

@article{wigner1993characteristic,
  title={Characteristic vectors of bordered matrices with infinite dimensions},
  author={Wigner, Eugene P},
  journal={The Collected Works of Eugene Paul Wigner: Part A: The Scientific Papers},
  pages={524--540},
  year={1993},
  publisher={Springer}
}

@article{casati1980connection,
  title={On the connection between quantization of nonintegrable systems and statistical theory of spectra},
  author={Casati, G and Valz-Gris, F and Guarnieri, I},
  journal={Lettere al Nuovo Cimento (1971-1985)},
  volume={28},
  pages={279--282},
  year={1980},
  publisher={Societ{\`a} Italiana di Fisica}
}

@article{brody1981random,
  title={Random-matrix physics: spectrum and strength fluctuations},
  author={Brody, T{\'o}mas A and Flores, Jorge and French, J Bruce and Mello, Pier A and Pandey, Akhilesh and Wong, Samuel SM},
  journal={Reviews of Modern Physics},
  volume={53},
  number={3},
  pages={385},
  year={1981},
  publisher={APS}
}

@article{guhr1998random,
  title={Random-matrix theories in quantum physics: common concepts},
  author={Guhr, Thomas and M{\"u}ller--Groeling, Axel and Weidenm{\"u}ller, Hans A},
  journal={Physics Reports},
  volume={299},
  number={4-6},
  pages={189--425},
  year={1998},
  publisher={Elsevier}
}

@article{berry1988semiclassical,
  title={Semiclassical formula for the number variance of the Riemann zeros},
  author={Berry, MV},
  journal={Nonlinearity},
  volume={1},
  number={3},
  pages={399},
  year={1988},
  publisher={IOP Publishing}
}

@article{rosenzweig1960repulsion,
  title={{``Repulsion of Energy Levels" in Complex Atomic Spectra}},
  author={Rosenzweig, Norbert and Porter, Charles E},
  journal={Physical Review},
  volume={120},
  number={5},
  pages={1698},
  year={1960},
  publisher={APS}
}

@article{vcadevz2024rosenzweig,
  title={{The Rosenzweig--Porter model revisited for the three Wigner--Dyson symmetry classes}},
  author={{\v{C}}ade{\v{z}}, Tilen and Nandy, Dillip Kumar and Rosa, Dario and Andreanov, Alexei and Dietz, Barbara},
  journal={New Journal of Physics},
  volume={26},
  number={8},
  pages={083018},
  year={2024},
  publisher={IOP Publishing}
}

@article{joyner2014gse,
  title={{GSE} statistics without spin},
  author={Joyner, Christopher H and M{\"u}ller, Sebastian and Sieber, Martin},
  journal={Europhysics Letters},
  volume={107},
  number={5},
  pages={50004},
  year={2014},
  publisher={IOP Publishing}
}

@article{rehemanjiang2016microwave,
  title={Microwave realization of the Gaussian symplectic ensemble},
  author={Rehemanjiang, Aimaiti and Allgaier, Markus and Joyner, CH and M{\"u}ller, S and Sieber, M and Kuhl, Ulrich and St{\"o}ckmann, H J},
  journal={Physical review letters},
  volume={117},
  number={6},
  pages={064101},
  year={2016},
  publisher={APS}
}

@article{akila2019gse,
  title={{GSE} spectra in uni-directional quantum systems},
  author={Akila, Maram and Gutkin, Boris},
  journal={Journal of Physics A: Mathematical and Theoretical},
  volume={52},
  number={23},
  pages={235201},
  year={2019},
  publisher={IOP Publishing}
}

@article{che2025experimental,
  title = {Experimental study of the distributions of off-diagonal scattering-matrix elements of quantum graphs with symplectic symmetry},
  author = {Che, Jiongning and Gluth, Nils and K\"ohnes, Simon and Guhr, Thomas and Dietz, Barbara},
  journal = {Phys. Rev. E},
  volume = {112},
  issue = {3},
  pages = {034208},
  numpages = {10},
  year = {2025},
  month = {Sep},
  publisher = {American Physical Society},
  doi = {10.1103/qm3x-jvyw},
}

@article{gomez2002misleading,
  title={Misleading signatures of quantum chaos},
  author={G{\'o}mez, JMG and Molina, Rafael A and Rela{\~n}o, Armando and Retamosa, Joaquin},
  journal={Physical Review E},
  volume={66},
  number={3},
  pages={036209},
  year={2002},
  publisher={APS}
}

@article{liu2018spectral,
  title={Spectral form factors and late time quantum chaos},
  author={Liu, Junyu},
  journal={Physical Review D},
  volume={98},
  number={8},
  pages={086026},
  year={2018},
  publisher={APS}
}

@article{prakash2021universal,
  title={Universal spectral form factor for many-body localization},
  author={Prakash, Abhishodh and Pixley, JH and Kulkarni, Manas},
  journal={Physical Review Research},
  volume={3},
  number={1},
  pages={L012019},
  year={2021},
  publisher={APS}
}

@article{das2025proposal,
  title={Proposal for many-body quantum chaos detection},
  author={Das, Adway Kumar and Cianci, Cameron and Cabral, Delmar GA and Zarate-Herrada, David A and Pinney, Patrick and Pilatowsky-Cameo, Sa{\'u}l and Matsoukas-Roubeas, Apollonas S and Batista, Victor S and del Campo, Adolfo and Torres-Herrera, E Jonathan and others},
  journal={Physical Review Research},
  volume={7},
  number={1},
  pages={013181},
  year={2025},
  publisher={APS}
}

@article{zarate2023generalized,
  title={Generalized survival probability},
  author={Zarate-Herrada, David A and Santos, Lea F and Torres-Herrera, E Jonathan},
  journal={Entropy},
  volume={25},
  number={2},
  pages={205},
  year={2023},
  publisher={MDPI}
}

@article{li2024spectral,
  title={Spectral form factor in chaotic, localized, and integrable open quantum many-body systems},
  author={Li, Jiachen and Yan, Stephen and Prosen, Toma{\v{z}} and Chan, Amos},
  journal={arXiv preprint arXiv:2405.01641},
  year={2024}
}

@article{daug2023many,
  title={Many-body quantum chaos in stroboscopically-driven cold atoms},
  author={Da{\u{g}}, Ceren B and Mistakidis, Simeon I and Chan, Amos and Sadeghpour, Hossein R},
  journal={Communications Physics},
  volume={6},
  number={1},
  pages={136},
  year={2023},
  publisher={Nature Publishing Group UK London}
}

@article{keating1997discrete,
  title={Discrete symmetries and spectral statistics},
  author={Keating, Jonathan P and Robbins, JM},
  journal={Journal of Physics A: Mathematical and General},
  volume={30},
  number={7},
  pages={L177--L181},
  year={1997}
}

@article{abanin2017recent,
 author = {Abanin, Dmitry A. and Papi\'c, Zlatko},
 title = {Recent progress in many-body localization},
 journal = {Ann. Phys.},
 volume = {529},
 number = {7},
 pages = {1700169},
 year = {2017},
 doi = {10.1002/andp.201700169},
}

@article{French1988,
title = {Statistical properties of many-particle spectra {VI. F}luctuation bounds on {N-NT}-noninvariance},
journal = {Ann. Phys.},
volume = {181},
number = {2},
pages = {235-260},
year = {1988},
issn = {0003-4916},
doi = {https://doi.org/10.1016/0003-4916(88)90166-2},
author = {J.B French and V.K.B Kota and A Pandey and S Tomsovic}
}

@article{Leviandier1956,
  title = {Fourier Transform: A Tool to Measure Statistical Level Properties in Very Complex Spectra},
  author = {Leviandier, Luc and Lombardi, Maurice and Jost, R\'emi and Pique, Jean Paul},
  journal = {Phys. Rev. Lett.},
  volume = {56},
  issue = {23},
  pages = {2449--2452},
  numpages = {0},
  year = {1986},
  month = {Jun},
  publisher = {American Physical Society},
  doi = {10.1103/PhysRevLett.56.2449},
  url = {https://link.aps.org/doi/10.1103/PhysRevLett.56.2449}
}

@Article{Kottos1999,
  Title                    = {Periodic orbit theory and spectral statistics for quantum graphs},
  Author                   = {Kottos, Tsampikos and Smilansky, Uzy},
  Journal                  = {Ann. Phys.},
  Year                     = {1999},
  Number                   = {1}, 
  Pages                    = {76--124},
  Volume                   = {274},
  Publisher                = {Elsevier}
}

@Article{Knill1998,
  Title                    = {On Nonconvex Caustics of Convex Billiards},
  Author                   = {Knill, Oliver},
  Journal                  = {Elemente der Mathematik},
  Year                     = {1998},
  Pages                    = {89},
  Volume                   = {53}
}

@article{Akila2015,
        doi = {10.1088/1751-8113/48/34/345101},
        year = 2015,
        month = {aug},
        publisher = {{IOP} Publishing},
        volume = {48},
        number = {34},
        pages = {345101},
        author = {Maram Akila and Boris Gutkin},
        title = {Spectral statistics of nearly unidirectional quantum graphs},
        journal = {J. Phys. A: Math. Theor.}
}

@Article{Gutkin2007,
  Title                    = {Dynamical "breaking" of time reversal symmetry},
  Author                   = {Boris Gutkin},
  Journal                  = {J. Phys. A},
  Year                     = {2007},
  Number                   = {31},
  Pages                    = {F761-F769},
  Volume                   = {40} 
}

@Article{Robbins1989,
  Title                    = {Discrete symmetries in periodic-orbit theory},
  Author                   = {Robbins, Jonathan M.},
  Journal                  = {Phys. Rev. A},
  Year                     = {1989},
  Month                    = {Aug},
  Pages                    = {2128--2136},
  Volume                   = {40}, 
  Doi                      = {10.1103/PhysRevA.40.2128},
  Issue                    = {4},
  Publisher                = {American Physical Society}
}

@article{Che2022,
  title = {Fluctuation properties of the eigenfrequencies and scattering matrix of closed and open unidirectional graphs with chaotic wave dynamics},
  author = {Che, Jiongning and Zhang, Xiaodong and Zhang, Weihua and Dietz, Barbara and Chai, Guozhi},
  journal = {Phys. Rev. E},
  volume = {106},
  issue = {1},
  pages = {014211},
  numpages = {14},
  year = {2022},
  month = {Jul},
  publisher = {American Physical Society},
  doi = {10.1103/PhysRevE.106.014211},
}

@Article{Dietz2014,
  Title                    = {Spectral properties and dynamical tunneling in constant-width billiards},
  Author                   = {Dietz, B. and Guhr, T. and Gutkin, B. and Miski-Oglu, M. and Richter, A.},
  Journal                  = {Phys. Rev. E},
  Year                     = {2014},
  Month                    = {Aug},
  Pages                    = {022903},
  Volume                   = {90},
  Doi                      = {10.1103/PhysRevE.90.022903},
  Issue                    = {2},
  Numpages                 = {15},
  Publisher                = {American Physical Society}
}

@article{abanin2019colloquium,
 author = {Abanin, Dmitry A. and Altman, Ehud and Bloch, Immanuel and Serbyn, Maksym},
 title = {Colloquium: Many-body localization, thermalization, and entanglement},
 journal = {Rev. Mod. Phys.},
 volume = {91},
 issue = {2},
 pages = {021001},
 numpages = {26},
 year = {2019},
 month = may,
 publisher = {American Physical Society},
 doi = {10.1103/RevModPhys.91.021001},
 url = {https://link.aps.org/doi/10.1103/RevModPhys.91.021001},
}

@article{Dietz2017,
  title = {Chaos and Regularity in the Doubly Magic Nucleus $^{208}\mathrm{Pb}$},
  author = {Dietz, B. and Heusler, A. and Maier, K. H. and Richter, A. and Brown, B. A.},
  journal = {Phys. Rev. Lett.},
  volume = {118},
  issue = {1},
  pages = {012501},
  numpages = {6},
  year = {2017},
  month = {Jan},
  publisher = {American Physical Society},
  doi = {10.1103/PhysRevLett.118.012501},
}

@article{Kolovsky2004a,
  title = {Quantum Chaos in the {{Bose-Hubbard}} Model},
  author = {Kolovsky, A. R. and Buchleitner, A.},
  year = {2004},
  month = nov,
  journal = {Europhysics Letters},
  volume = {68},
  number = {5},
  pages = {632},
  publisher = {IOP Publishing},
  issn = {0295-5075},
  doi = {10.1209/epl/i2004-10265-7},
  urldate = {2023-03-15},
  langid = {english}
}

@book{Cohen-Tannoudji2011a,
  title = {Advances in {{Atomic Physics}}: {{An Overview}}},
  shorttitle = {Advances in {{Atomic Physics}}},
  author = {{Cohen-Tannoudji}, Claude and {Gu{\'e}ry-Odelin}, David},
  year = {2011},
  month = sep,
  publisher = {WORLD SCIENTIFIC},
  doi = {10.1142/6631},
  urldate = {2025-09-29},
  isbn = {978-981-277-496-5 978-981-277-498-9},
  langid = {english}
}

@article{Berry1977,
  title = {Level Clustering in the Regular Spectrum},
  author = {Berry, Michael Victor and Tabor, M.},
  year = {1977},
  month = jan,
  journal = {Proceedings of the Royal Society of London. A. Mathematical and Physical Sciences},
  volume = {356},
  number = {1686},
  pages = {375--394},
  publisher = {Royal Society},
  doi = {10.1098/rspa.1977.0140},
  urldate = {2025-10-07}
}

@article{Chakrabarti2003,
  title = {Level Correlation in Coupled Harmonic Oscillator Systems},
  author = {Chakrabarti, Barnali and Hu, Bambi},
  year = {2003},
  month = aug,
  journal = {Physics Letters A},
  volume = {315},
  number = {1},
  pages = {93--100},
  issn = {0375-9601},
  doi = {10.1016/S0375-9601(03)01001-6},
  urldate = {2025-09-12}
}

@book{Stockmann1999a,
  title = {Quantum {{Chaos}}: {{An Introduction}}},
  shorttitle = {Quantum {{Chaos}}},
  author = {St{\"o}ckmann, Hans J{\"u}rgen},
  year = {1999},
  month = oct,
  edition = {1},
  publisher = {Cambridge University Press},
  doi = {10.1017/CBO9780511524622},
  urldate = {2025-10-07},
  copyright = {https://www.cambridge.org/core/terms},
  isbn = {978-0-521-59284-0 978-0-521-02715-1 978-0-511-52462-2}
}

@article{Chakrabarti2012,
  title = {Energy-Level Statistics of Interacting Trapped Bosons},
  author = {Chakrabarti, Barnali and Biswas, Anindya and Kota, V. K. B. and Roy, Kamalika and Haldar, Sudip Kumar},
  year = {2012},
  month = jul,
  journal = {Physical Review A},
  volume = {86},
  number = {1},
  pages = {013637},
  issn = {1050-2947, 1094-1622},
  doi = {10.1103/PhysRevA.86.013637},
  urldate = {2025-01-28},
  copyright = {http://link.aps.org/licenses/aps-default-license},
  langid = {english}
}

@article{Roy2012,
  title = {Spectral Fluctuation and $1/f^\alpha$ Noise in the Energy Level Statistics of Interacting Trapped Bosons},
  author = {Roy, Kamalika and Chakrabarti, Barnali and Biswas, Anindya and Kota, V. K. B. and Haldar, Sudip Kumar},
  year = {2012},
  month = jun,
  journal = {Physical Review E},
  volume = {85},
  number = {6},
  pages = {061119},
  publisher = {American Physical Society},
  doi = {10.1103/PhysRevE.85.061119},
  urldate = {2025-10-07}
}

@article{Werner2012,
  title = {General Relations for Quantum Gases in Two and Three Dimensions: {{Two-component}} Fermions},
  author = {Werner, F{\'e}lix and Castin, Yvan},
  year = {2012},
  month = jul,
  journal = {Physical Review A},
  volume = {86},
  number = {1},
  pages = {013626},
  publisher = {American Physical Society},
  issn = {1050-2947},
  doi = {10.1103/PhysRevA.86.013626},
  urldate = {2017-05-15}
}

@article{Gross2017,
  title = {Quantum Simulations with Ultracold Atoms in Optical Lattices},
  author = {Gross, Christian and Bloch, Immanuel},
  year = {2017},
  month = sep,
  journal = {Science},
  volume = {357},
  number = {6355},
  pages = {995--1001},
  issn = {0036-8075},
  doi = {10.1126/science.aal3837},
  pmid = {28883070}
}

@article{Schafer2020,
  title = {Tools for Quantum Simulation with Ultracold Atoms in Optical Lattices},
  author = {Sch{\"a}fer, Florian and Fukuhara, Takeshi and Sugawa, Seiji and Takasu, Yosuke and Takahashi, Yoshiro},
  year = {2020},
  month = aug,
  journal = {Nature Reviews Physics},
  volume = {2},
  number = {8},
  pages = {411--425},
  publisher = {Nature Publishing Group},
  issn = {2522-5820},
  doi = {10.1038/s42254-020-0195-3},
  urldate = {2023-04-13},
  copyright = {2020 Springer Nature Limited},
  langid = {english}
}

@unpublished{Bradly2025,
    author = {Bradly, C. J. and Brand, J.},
    title = {Prospects for creating a metastable {Laughlinian} few-boson gas with cold atoms},
    date = {2025},
    note = {in preparation}
}

\end{document}